\documentclass[aps,prl,twocolumn]{revtex4-2}

\usepackage{graphicx} 
\usepackage{array}
\usepackage[usenames,dvipsnames]{xcolor}

\definecolor{cream}{RGB}{222,217,201}

\graphicspath{{figs/}}

\usepackage{subcaption}
\usepackage{nicefrac} 
\usepackage{url}
\usepackage{amsmath}

\usepackage{xargs}
\usepackage{color}
\usepackage{mdframed}

\usepackage[color]{changebar}
\cbcolor{red}

\newcommand{\rvec}{\mathbf{r}}
\newcommand{\kvec}{\mathbf{k}}

\newcommand\V{\Phi}
\newcommand\Vk{\tilde\Phi(\kvec)}
\newcommand\volume{V}

\newcommand\pfull{\ensuremath{p_{\text{full}}}}
\newcommand\pempty{\ensuremath{p_{\text{empty}}}}

\newcommand\mufull{\ensuremath{\mu_{\text{full}}}}
\newcommand\muempty{\ensuremath{\mu_{\text{empty}}}}

\newcommand\rhofull{\ensuremath{\rho_{\text{full}}}}
\newcommand\rhoempty{\ensuremath{\rho_{\text{empty}}}}

\newcommand\gst{\ensuremath{\Delta g_{st}}}

\definecolor{cream}{RGB}{222,217,201}

\begin{document}

\author{Jordan Pommerenck}
\affiliation{Department of Physics, Oregon State University, Corvallis, OR 97331}
\author{Cory Simon}
\affiliation{Department of Chemical Engineering, Oregon State University, Corvallis, OR 97331}
\author{David Roundy}
\affiliation{Department of Physics, Oregon State University, Corvallis, OR 97331}

\title{An upper bound to gas storage and delivery via pressure-swing adsorption in
porous materials}

\begin{abstract}
    Both hydrogen
    and natural gas are challenging to economically store onboard vehicles as
    fuels, due to their low volumetric energy density at ambient conditions.
    One strategy to densify
    these gases is to pack the fuel tank with a porous
    adsorbent material.
    The US Department of
    Energy (DOE) has set volumetric deliverable capacity
    targets which, if met, would help
    enable commercial adoption of hydrogen/natural gas as transportation fuels.
    %
    %
    Here, we present a
    theoretical upper bound on the deliverable capacity of a gas in a rigid
    porous
    material via an isothermal pressure swing.
    To provide an extremum, we consider
    a substrate that provides a spatially uniform potential energy field for the
    gas. Our bound relies directly on experimentally measured
    properties of the pure gas.
    We conclude
    that the deliverable capacity targets set by the DOE for room-temperature natural gas and hydrogen
    storage are just barely theoretically possible.
    The targets are likely to be impossible for any real,
    rigid porous material because of steric
    repulsion, which reduces the deliverable capacity below our upper
    bound. Limitations
    to the scope of applicability of our upper bound may guide fuel tank design and
    future material development. Firstly, one could avoid using an \emph{isothermal}
    pressure swing by heating the adsorbent to drive off
    trapped, residual gas. Secondly,
    our upper bound assumes the material does not change its
    structure in response to adsorbed gas, suggesting that flexible materials could
    still satisfy the DOE targets.    
\end{abstract}

\maketitle


\section{Introduction}
The transportation sector is dominantly powered by petroleum-based
fuels~\cite{davis2009transportation}. 
Consequently, in the United States, it is responsible
for 36\% of energy-related carbon dioxide emissions \cite{useia} and $\sim$50,000
premature deaths per year associated with particulate matter and ozone
emissions~\cite{caiazzo2013air}. 
Natural gas and hydrogen (H$_2$) are
alternative transportation fuels that, if adopted widely,
could mitigate climate change~\cite{mcglade2015geographical} and improve air quality for human
health. 
More, the finite global petroleum resources are declining
rapidly~\cite{sorrell2010global}.
The development of technologies for the
widespread adoption of sustainable transportation fuels, such as hydrogen, is therefore
critical.

Natural gas, mostly methane, is considered a transition (to a renewable and
clean) fuel because it emits 25\% less carbon dioxide~\cite{eia2013much} and
fewer toxic byproducts~\cite{wang2000full} upon combustion per unit energy
produced compared to gasoline. From an economic standpoint, the supply of
natural gas in the United States is increasing as a result of hydraulic
fracturing and horizontal drilling techniques~\cite{usnatgassupply}. A positive
environmental outlook for natural gas, however, is predicated on mitigating
fugitive emissions (methane is itself a potent greenhouse
gas)~\cite{alvarez2012greater} and groundwater
contamination~\cite{osborn2011methane} from hydraulic fracturing.

Hydrogen (H$_2$) is the ultimate transportation fuel because it emits only
water when it electrochemically reacts with oxygen in a fuel cell to power a
vehicle. 
Notably, the environmental allure of hydrogen is predicated on its production 
via a renewable means, e.g.\ splitting water using wind-generated electricity;
currently, hydrogen (H$_2$) is primarily produced by steam reforming
of natural gas followed by the water-gas shift reaction, which emits carbon
dioxide \cite{crabtree2004hydrogen}. 


At ambient conditions, both methane and hydrogen gas possess a very low
volumetric energy density compared to (liquid) gasoline. Consequently, under storage
space constraints in passenger vehicles, natural gas and hydrogen must be
densified for onboard storage to achieve a reasonable driving range on a
``full'' tank of fuel. Traditional densification approaches are liquefaction,
at cryogenic temperatures and atmospheric pressure, or compression, at room
temperature and high pressures. Both approaches require expensive
infrastructure at refilling stations and significant energy input; e.g., the
energy input to liquify hydrogen is $\sim$30\% of its energy
content~\cite{bossel2003energy}. Also, high-pressure storage tanks are
heavy, thick-walled, and non-conformable, while cryogenic storage tanks are
bulky, expensive, and afflicted by boil-off losses~\cite{hasan2009minimizing}.

A promising approach to densify natural
gas~\cite{makal2012methane,mason2014evaluating} and
hydrogen~\cite{suh2011hydrogen,garcia2018benchmark} for vehicular storage at
room temperature is through physical adsorption in nanoporous
materials~\cite{schoedel2016role}. The internal surfaces of porous materials
attract gas molecules through van der Waals, electrostatic, etc.\ interactions
to achieve a higher adsorbed gas density than the bulk gas at the same
temperature and pressure.
Nanoporous materials could thus allow for room temperature and lower-pressure
storage of natural gas and hydrogen and alleviate many drawbacks of high-pressure or low-temperature
storage.

\begin{figure}
    \centering
    \includegraphics[width=0.8\columnwidth]{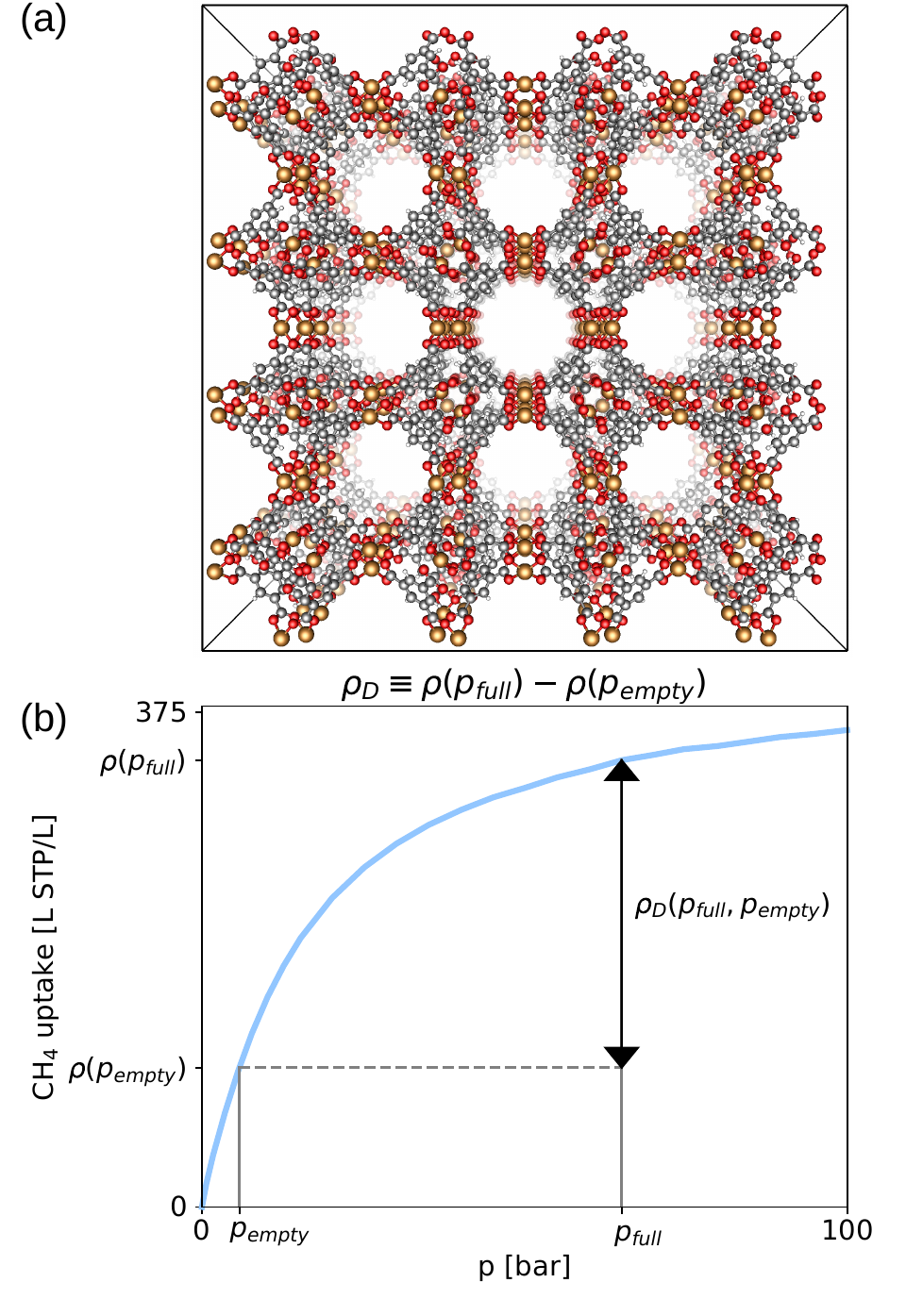}
    \begin{minipage}{0.25\textwidth}
    \phantomsubcaption{\label{fig:example_MOF}}
    \end{minipage}
    \begin{minipage}{0.25\textwidth}
    \phantomsubcaption{\label{fig:delcap}}
    \end{minipage}
    
    \caption{\label{fig:fig1} Gas storage and delivery using metal-organic frameworks (MOFs). (\subref{fig:example_MOF}) the crystal structure of an archetype MOF, CuBTC \cite{chui1999chemically}. (\subref{fig:delcap}) the methane adsorption isotherm in CuBTC \cite{chui1999chemically} at 298 K (blue) (data from Ref.~\cite{mason2014evaluating}). The deliverable capacity $\rho_D$ is illustrated as the density of gas in the MOF at the storage pressure $\pfull$ minus the density at the discharge pressure $\pempty$.}
\end{figure}

For a vehicle employing a porous material to store natural gas or hydrogen, the
(volumetric) \emph{deliverable capacity} of the gas is the primary
thermodynamic property that determines the driving
range on a given ``full'' tank of fuel~\cite{mason2014evaluating}. 
The adsorbed gas storage tank delivers the
gaseous fuel to the engine via an (assumed) isothermal pressure
swing~\cite{sircar2002pressure}. The deliverable capacity (see
Fig.~\ref{fig:delcap}) is the density of the gas in the material at the storage
pressure $\pfull$ minus the residual gas that remains adsorbed at the lowest
pressure $\pempty$ such that sufficient flow is maintained to feed the engine.
For commercial feasibility, the US Department of Energy (DOE) has set
deliverable capacity targets for both adsorbed methane and hydrogen storage and
delivery for vehicles. For methane, the Advanced Research Projects
Agency--Energy (ARPA-E) set a deliverable capacity target of 315\ L STP CH$_4$
per L of adsorbent at 298 K using a 65\,bar to 5.8\,bar pressure
swing~\cite{simon2015materials}. For hydrogen, the DOE set a series of
progressive targets at five year intervals, with the ultimate target of 0.05\,kg
H$_2$/L~\cite{h2targetsDOE} using a 100\ bar to 5\ bar pressure swing at a
minimum of -40 $^\circ$C~\cite{allendorf2018assessment}. Thus far, despite the
emergence of highly tunable materials with large surface areas, such as
metal-organic frameworks~\cite{furukawa2013chemistry} and covalent organic 
frameworks~\cite{diercks2017atom}, no porous material has met these deliverable capacity targets.


To set realistic performance targets and optimally allocate resources to research efforts,
in this work, we present a theoretical framework that places an intrinsic upper
limit on the deliverable capacity of a pure gas in a rigid porous material and
uses as input the experimentally measured properties of the bulk gas. Our
extremum is provided by a substrate that offers a spatially uniform potential
energy field felt by the gas. Applying our framework to methane and hydrogen
gas, we find the US DOE deliverable capacity targets for natural gas and
hydrogen storage and delivery are theoretically possible, but sufficiently
close to the upper bound as to be impractical for any real, rigid porous
material. Optimistically, new paradigms outside the scope of applicability of
our theoretical framework, such as gas-induced structural transitions of the
material, hold promise for meeting these targets, as evidenced by flexible MOF
Co(bdp) which currently boasts the largest methane deliverable
capacity~\cite{mason2015methane}.

\section{Gas storage \& delivery by isothermal, pressure-swing adsorption}
Consider a pressure vessel onboard a vehicle (i.e. fuel tank) packed with
porous material. At the filling stage, the tank is connected to a (pure)
gaseous reservoir at pressure $\pfull$ and allowed to equilibrate. At this
point, the adsorbed gas tank is considered full. While driving, gas desorbs
from the adsorbent to the engine/fuel cell, driven by a pressure differential.
The tank is considered depleted/empty when the pressure has dropped to
$\pempty$, the pressure at which the flow rate of gas from the tank to the
engine is insufficient. However, given $\pempty \neq 0$ (pulling vacuum),
residual gas will remain trapped in the adsorbent. Therefore, the driving range
of the vehicle is primarily determined by the \emph{deliverable capacity} $\rho_D$ of
the gas in the material (see Fig.~\ref{fig:delcap}): the density of gas at
$\pfull$ minus that at $\pempty$. The isothermal, volumetric deliverable
capacity is an intrinsic property of the nanoporous material and its
interaction with the gas.

\section{Review of previous work}
There has been considerable work attempting to establish an upper bound on the
isothermal deliverable capacity in pressure-swing adsorption.

Early work showed that, in the simplified Langmuir model, there exists an
optimal free energy of adsorption (which determines the Langmuir constant) that
maximizes the deliverable capacity~\cite{matranga1992storage,bhatia2006optimum,simon2014optimizing}. If
the gas-substrate interaction is too weak (strong), too little (much) gas
adsorbs (is retained) at $\pfull$ ($\pempty$), diminishing the deliverable capacity. An upper
bound on the deliverable capacity of a Langmuir material follows if each
adsorption site is endowed with the optimal free energy of adsorption. However,
remaining is the question of how many adsorption sites per volume a porous
material can practically offer, under the constraint that these adsorption
sites provide the optimal free energy of adsorption. Moreover, gas-gas
attractions, neglected in the Langmuir model, could recruit more gas in the
material at $\pfull$ than at $\pempty$ and enhance the deliverable
capacity~\cite{simon2014optimizing}.

G\'omez-Gualdr\'on \emph{et al.}~\cite{gomez2017impact} introduced a model that
accounted for gas-gas interactions via an intermolecular potential and
idealized the substrate in two different ways---as a: (1) set of discrete adsorption sites
packed into an FCC lattice and (2) a volume endowed with a spatially uniform,
background potential energy field. According to molecular simulation of methane
adsorption, the ARPA-E deliverable capacity target of 315 L STP/L could be
reached in both of these idealized substrates. However, both models (i)
neglect the space occupied by atoms of the porous material that are needed to
endow the adsorption sites/volume with the attractive potential energy and (ii)
rely on a molecular model for methane.

A third body of work incorporated both gas-gas interactions \emph{and} steric
interactions of the gas with the atoms of the porous material. Simulations of
methane adsorption in hundreds of thousands of explicit nanoporous crystal
structures---both real and hypothetical---suggested that the ARPA-E deliverable
capacity target is infeasible~\cite{simon2015materials}; the highest simulated
methane deliverable capacity was 196 cm$^3$~STP/cm$^3$. Confidence in this
conclusion rests upon (i) the accuracy of the intermolecular potentials
describing the molecular interactions and (ii) the sufficient sampling of
material space, i.e.\ that the structures considered are representative of the
set of possible materials \cite{Moosavi2020understanding}. To further explore material space and address
sensitivity to the intermolecular potentials: scaling the Lennard-Jones
potential well depths of material atoms to model enhanced
interactions~\cite{gomez2014exploring}, placing Lennard-Jones spheres in a unit
cell randomly to form ``pseudo-materials''~\cite{kaija2018high}, and generating
fictitious potential energy fields via a generative adversarial network trained
on zeolite structures~\cite{lee2019predicting} all generated model substrates
that failed to meet the ARPA-E methane deliverable capacity target.

Finally, in a number of studies, molecular simulations of gas adsorption in
a large pool of both existing and predicted nanoporous
structures empirically shed light on performance limitations \cite{firlej2013understanding, goldsmith2013theoretical,
ahmed2019exceptional,simon2015materials}.
Such studies can guide the development of new materials
by allowing researchers to identify commonalities among materials that have
desirable properties.  
However, the largest observed simulated deliverable capacity among the pool of candidate materials in such studies
is only a \emph{lower bound} on the highest attainable deliverable capacity that is is predicted upon the accuracy of the molecular models.


In this work, we place a rigorous upper bound on the isothermal deliverable capacity of a
pure gas in a rigid substrate by endowing a control volume with a spatially
uniform background energy field. Instead of using a molecular model for the
gas~\cite{gomez2017impact}, we use the experimental equation of state to
account for gas-gas interactions. In addition, we prove using the calculus of
variations that this spatially uniform substrate yields a maximal deliverable
capacity. We use our framework to place an upper bound on the deliverable
capacity of methane and hydrogen gas at conditions relevant to vehicular storage.

\section{Methods}
\subsection{An upper bound on the deliverable capacity}\label{sec:upper-bound}
We now develop a thermodynamic framework that places an upper bound on the deliverable capacity of any pure gas in a rigid porous material.

\begin{figure}
    \centering
    \includegraphics[width=\columnwidth]{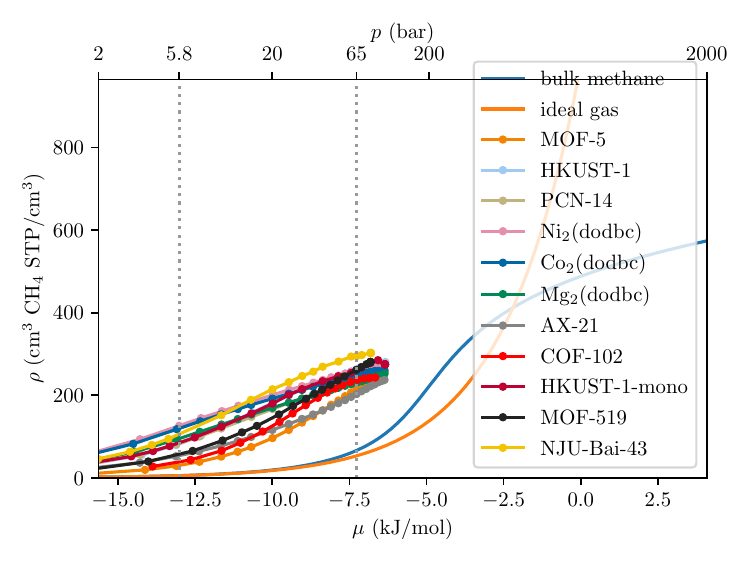}
    \caption{The density of bulk methane, the ideal gas, and adsorbed gas in several porous materials at 298\,K as a function of chemical potential (bottom axis) and pressure (top axis). The bulk methane density is from the National Institute of Standards and Technology (NIST)~\cite{nist}. The methane adsorption isotherms in the porous materials are experimental data from Ref.~\cite{mason2014evaluating, furukawa2009storage, tian2018sol, gandara2014high, zhang2017fine}.
    }
    \label{fig:density-vs-mu-ch4}
  \end{figure}

The thermodynamic properties of a bulk, pure gas are characterized by an
equation of state. Of particular interest for gas storage and delivery is the
density of the gas $\rho_g(\mu,T)$ as a function of chemical potential $\mu$
and temperature $T$. We show $\rho_g(\mu, T)$ for methane gas in
Fig.~\ref{fig:density-vs-mu-ch4} at $T=298$\ K. For comparison, we also show
the density of methane adsorbed into several porous materials.

To place an upper bound on the deliverable capacity, consider a substrate whose
sole interaction with the gas is to introduce a spatially uniform potential
energy $\V$ for gas molecules in the control volume (where $\V<0$ for an
attractive potential). Because this interaction is uniform within the
substrate, the gas-gas interactions and thus fluid structure in such a
substrate are identical to those of the pure gas at the same density and
temperature. This allows us to obtain the adsorption properties of this model
substrate using the experimentally measured properties of the pure
gas~\cite{nist}. Intuitively, the deliverable capacity of a gas in such a
homogeneous substrate with the optimal potential energy is an upper bound
because, in a real material, (i) spatial inhomogeneity of the potential
energy results in some points in the control volume offering a suboptimal
attraction for the gas and (ii) the atoms required to create the potential
field exclude gas from occupying a fraction of the control volume and thus reduce utilized space.

The spatially uniform potential $\V$ representing gas-substrate interactions
behaves as an external potential and effectively shifts the chemical potential
(or, equivalently, molar Gibbs free energy) of the gas in the material, just as
gravitational potential energy causes the density of air to vary with altitude.
Consequently, the density of gas in our homogeneous substrate is:
\begin{align}
    \rho(\mu,T) &= \rho_g(\mu - \V,T). \label{eq:mof-density}
\end{align}
See Sec.~\ref{sec:V_shifts_chem_pot} for a derivation.

\begin{figure}
    \centering
    \includegraphics[width=0.95\columnwidth]{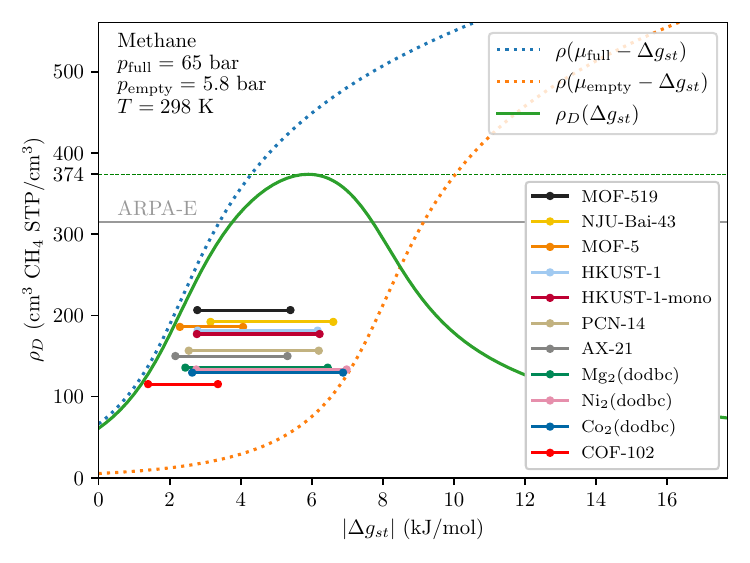}
    \caption{The deliverable capacity of methane as a function of the attractive Gibbs free energy $|\gst|$.
    Experimental deliverable capacities for several porous materials (data from Ref.~\cite{mason2014evaluating, furukawa2009storage, tian2018sol, gandara2014high, zhang2017fine}) are shown along with the experimental values for $\gst$ at the empty and full pressures shown as dots connected by a line.}
    \label{fig:methane-298-D}
\end{figure}

The deliverable capacity of gas in our homogeneous substrate is thus:
\begin{align}
    \rho_D(\V,T) &= \rho(\mufull,T) - \rho(\muempty,T),
    \label{eq:DofPhi}
    \\
    &= \rho_g(\mufull-\V,T) - \rho_g(\muempty-\V,T),
\end{align}
where $\mufull$ and $\muempty$ are the chemical potentials corresponding to the
pressures $\pfull$ and $\pempty$, respectively. Thus, $\rho_D$ for our
homogeneous substrate as a function of $\V$ is the difference between two
shifted versions of $\rho_g(\mu; T)$. Figure~\ref{fig:methane-298-D} shows the
two shifted bulk methane density curves and their difference. The horizontal
axis is the difference in molar Gibbs free energy of the gas, at fixed density (the density in the adsorbent) and temperature, between the gas within the substrate and the pure gas (\gst),
which is equal to $\V$ in our model, but is pressure-dependent in a real
crystal~(see Sec.~\ref{sec:phi-is-delta-g}). We see a potential $\V$ that
maximizes the deliverable capacity of our ideal substrate, balancing the need
to maximize the density at $\pfull$ against the need to minimize residual gas
retained at $\pempty$.

The optimal deliverable capacity of methane (298\ K, $\pfull=$ 65\ bar,
$\pempty=$ 5.8\ bar) in our ideal, spatially uniform substrate is 374\ L STP/L,
achieved for $\V_{opt} =$ 5.9\ kJ/mol.

An essential question is whether our idealized substrate, with spatially
uniform potential $\V_{opt}$ for the gas, places an upper bound on the
deliverable capacity in all rigid porous materials. In
Section~\ref{sec:proof-extremum}, we show that our idealized substrate with
spatially uniform potential $\V_{opt}$ yields an \emph{extremum} of the
deliverable capacity over all possible \emph{static potential energy fields},
provided the gas does not crystallize at the temperature and in the density
range of interest. The potential energy field is \emph{static} if it is
unaffected by the presence of the gas; consequently, our model does not apply
to flexible materials that undergo gas-induced conformation
changes~\cite{schneemann2014flexible}. We next argue that this extremum is a
maximum by addressing two alternative possibilities: the extremum could turn
out to be (i) a saddle point or (ii) a local, rather than global, maximum.

Fig.~\ref{fig:methane-298-D} shows that $\V_{opt}$ provides a maximum
$\rho_D$ over all \emph{spatially uniform}, static potential energy fields. To
qualitatively argue that the spatially uniform potential $\V_{opt}$ provides a
maximum deliverable capacity over \emph{all} (including non-spatially uniform)
static potential energy fields (a broader claim), consider a
non-spatially-uniform variation on $\V_{opt}$. 
At low adsorbed gas densities, the lowest-energy positions will be preferentially occupied, while
at higher densities, the additional gas molecules will be forced into higher
energy locations. Thus, the mean attraction will be lowest at \pfull\ and
highest at \pempty. 
Therefore, the effect of a non-spatially uniform variation on $\V_{opt}$ is to reduce the deliverable capacity.
By the same argument, $\gst$ increases monotonically with
pressure. As shown in Sec.~\ref{sec:monotonic}, this results in
the deliverable capacity provided by a non-uniform potential with a given $\gst$ at $\pfull$ or $\pempty$ being lower than the
deliverable capacity provided by a uniform potential with $\V=\gst$. Thus, a spatially uniform potential gives the maximal
deliverable capacity amongst all static potential energy fields that give the
same $\gst$, and the \emph{optimal} uniform potential $\V_{opt}$ will yield a
greater deliverable capacity over any static, non-uniform potential.

\subsection{An external, spatially uniform potential $\V$ shifts the chemical potential $\mu$ of the gas} \label{sec:V_shifts_chem_pot}
We show that imposing an external, spatially uniform potential $\V$ to a gas
has the effect of shifting the chemical potential $\mu$ of the gas. Consider a control volume $\Omega$ with volume
$V=|\Omega|$ (large enough to neglect boundary effects) that is endowed with
the external, spatially uniform potential $\V$. Impose the grand-canonical
ensemble, where this control volume can exchange energy and particles with a
bath of gas at temperature $T$ and chemical potential $\mu$.

To denote a microstate of this system, let $N$ be the number of gas particles
in the control volume and $\mathbf{r}_1,...,\mathbf{r}_N$ be their positions.
Then, the potential energy $E$ of a microstate of the control volume is:

\begin{equation} E(\rvec_1,...,\rvec_N) = N\V +
E_{gg}(\rvec_1,...,\rvec_N),
\end{equation}
where $E_{gg}$ is the (unknown and complicated) interatomic potential for
gas-gas interactions that governs the (real) gas properties. The first term
arises from each gas molecule experiencing the external potential $\V$, where
$\V < 0$ corresponds to attraction. To account for molecular rotational and
vibrational degrees of freedom, we could treat the positions $\rvec_i$ as the
locations of atoms rather than molecules. In this case the gas-gas interactions
will include both \emph{intra-molecular} interactions and
\emph{inter-molecular} interactions.

The grand canonical partition function of the control volume is:
\begin{multline}
    \Xi(\mu, V, T)= \\ \displaystyle \sum_{N=0}^\infty \frac{1}{\Lambda^{3N}N!} \int_{\Omega} \cdots \int_{\Omega} e^{-\beta E_{gg}(\rvec_1, ..., \rvec_N)} e^{\beta (\mu - \V) N} d\rvec_1 \cdots d\rvec_N.
    \label{eq:gcpf}
\end{multline}
We recognize this as the grand canonical partition function of the bulk gas in
the control volume without the external potential, but shifted in chemical
potential:
\begin{equation}
    \Xi(\mu, V, T)=\Xi_g(\mu - \V, V, T)
    \label{eq:xi_vs_xi0}
\end{equation}
where $\Xi_g(\mu, V, T)$ is the grand partition function of the bulk
gas in the control volume in the absence of an external potential. Importantly,
this equivalency depends on the potential $\V$ being spatially uniform.
Therefore, the thermodynamic properties of the gas atoms in the spatially
uniform external potential $\V$ (adsorbed in our idealized substrate)
are equivalent to the properties of the bulk gas at chemical potential $\mu-\V$
(where $T$ is held fixed). As $\V$ becomes more negative, corresponding to a
more attractive adsorbent, the thermodynamic properties of the adsorbed gas in
our ideal substrate are equivalent to the gas at a higher chemical potential.

\subsection{Proof of extremum}\label{sec:proof-extremum}

To show that a uniform potential gives the highest deliverable capacity, we
consider an interaction potential between gas and substrate offering a
potential energy field $\V(\rvec)$ that varies in space. We will show that the
uniform potential has the highest deliverable capacity of the set of potentials
having the same mean value, and to find the maximum deliverable capacity one
maximizes over this mean value. In this proof, we make use of the Fourier
transform of this potential:
\begin{align}
    \Vk \equiv \iiint \V(\rvec) e^{-i\kvec\cdot \rvec} d\rvec
\end{align}
The Fourier transform of a uniform potential is a Dirac delta function
$\tilde{\V}(\kvec)\propto\delta(\kvec)$. For a uniform potential to extremize
the deliverable capacity, we must show that the functional derivative of the
deliverable capacity with respect to $\Vk$ is zero for \emph{nonzero} values of
$\kvec$, i.e.
\begin{align}
    \frac{\delta \rho_D}{\delta \Vk} &= \mathbf{0}, \text{ if } \kvec\ne \mathbf{0}.
\end{align}
We note that this functional derivative may be non-zero for $\kvec=\mathbf{0}$
because we separately maximize with respect to the mean value of the uniform
potential $\V$. This means that
\begin{align}
    \frac{\delta N_\text{full}}{\delta \Vk} &= \frac{\delta N_\text{empty}}{\delta \Vk}
\end{align}
where $N_\text{full}$ and $N_\text{empty}$ are the number of particles in the
substrate at the full and empty pressure, $\pfull$ and $\pempty$, respectively.

Because the chemical potential $\mu$ varies monotonically with $N$ at fixed
temperature, we can consider how the chemical potential varies as we change
$\Vk$ with the number of molecules held fixed. We demonstrate this using the
cyclic chain rule, which shows us that
\begin{align}
    \left(\frac{\delta N}{\delta \Vk}\right)_{\mu} &=
    -\left(\frac{\delta \mu}{\delta \Vk}\right)_{N}
    \left(\frac{\partial N}{\partial \mu}\right)_{\Vk}.
\end{align}
Since changing the chemical potential changes the number of molecules in the
general case, if we can show that $\left(\frac{\delta \mu}{\delta
\Vk}\right)_{N}=0$ then we will have shown that $\left(\frac{\delta N}{\delta
\Vk}\right)_{\mu}=0$. Thus we consider
\begin{align}
    \left(\frac{\delta \mu}{\delta \Vk}\right)_N
    &= \left(\frac{\delta \left(\frac{\partial F}{\partial N}\right)_{\volume}}{\delta \Vk}\right)_N
    \\
    &= \left(\frac{\partial \left(\frac{\delta F}{\delta \Vk}\right)_{N}}{\partial N}\right)_{\volume}
    \label{eq:dmudpot}
\end{align}
where we have made use of the derivative relationship between $\mu$ and the
Helmholtz free energy $F$, and have then reordered the functional and partial
derivatives. Let us consider the interior derivative first. The derivative of
the Helmholtz free energy with respect to the external potential $\Vk$ yields
the number density:
\begin{align}
    \frac{\delta F}{\delta \Vk} &= \rho(\kvec)
\end{align}
The number density given a uniform potential is itself spatially uniform for
any stable system in a fluid state at this density (i.e. does not spontaneously
crystallize), and thus has a Fourier transform that is proportional to a Dirac
$\delta$-function. Thus, the functional derivative $\frac{\delta F}{\delta
\V(\rvec)}$ is actually a uniform function. We can insert this expression into
Eq.~\ref{eq:dmudpot} to find that
\begin{align}
    \left(\frac{\delta \mu}{\delta \Vk}\right)_N &\propto \delta(\kvec) \\
    \left(\frac{\delta N}{\delta \Vk}\right)_\mu &\propto \delta(\kvec)
\end{align}
Thus, the functional derivative of both $\mu$ and $N$ with regard to
$\V(\rvec)$ are themselves spatially uniform. Since we already maximize
$\rho_D$ with respect to the spatially uniform component of the potential (i.e.
$\kvec=0$), the derivative of $\rho_D$ with respect to any change of potential
is zero.

This demonstrates that a spatially uniform potential leads to an extremum value
of the deliverable capacity. This proof is insufficient, however, to show that
it must be a true maximum.

\subsection{The real-substrate analog of $\V$ is $\gst$}
\label{sec:phi-is-delta-g}
\begin{figure}
    \centering
    \includegraphics[width=\columnwidth]{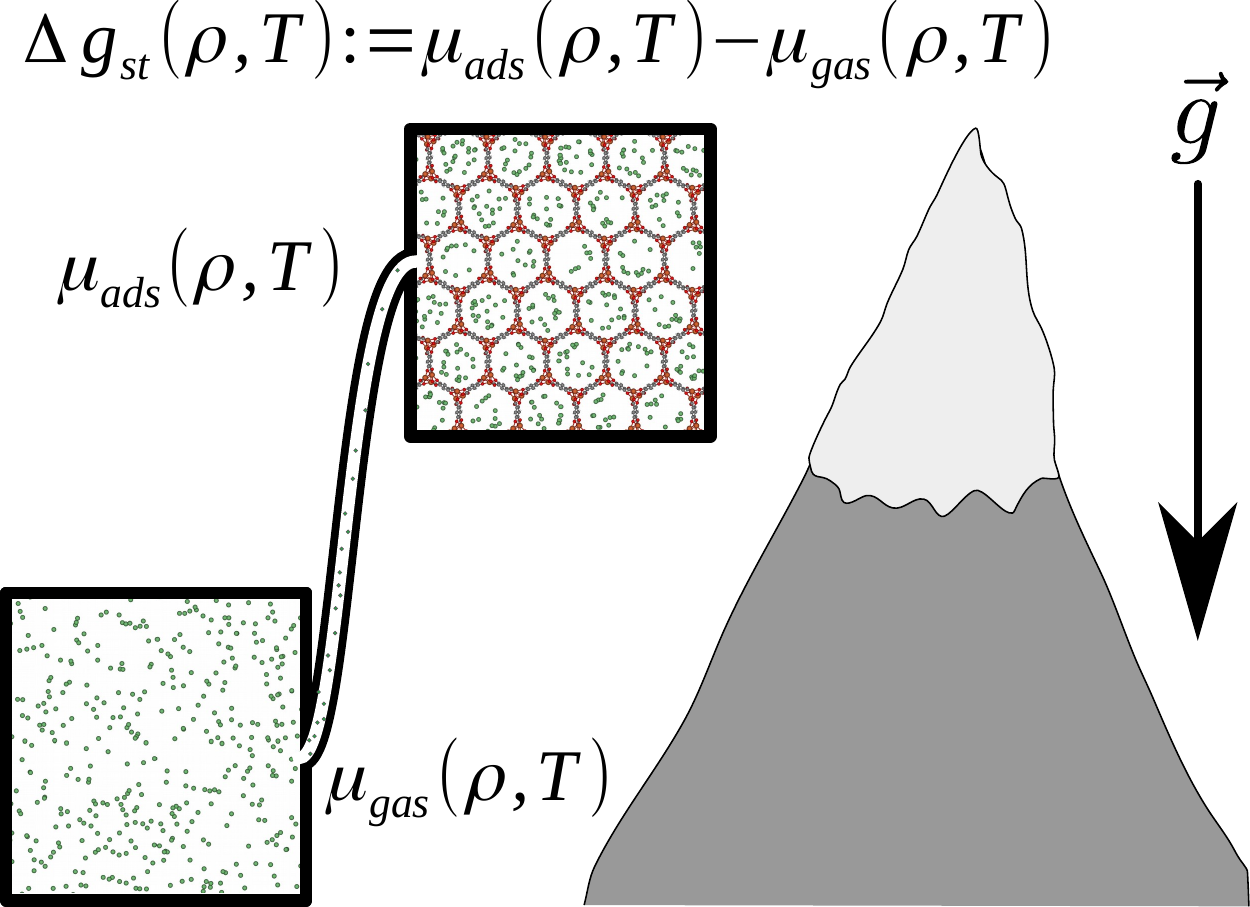}
    \caption{An illustrative thought experiment to measure $\gst$ by adjusting the altitute difference between two connected volumes, one free space and another containing a porous material.
    $\gst$ is the difference in the gravitational potential energy when the density of gas in each volume is equal.
    }
    \label{fig:delta-G-cartoon}
\end{figure}

The parameter describing our idealized substrate is $\V$, the spatially uniform
potential felt by a gas molecule adsorbed in the idealized substrate. A natural
question is how this potential relates to the properties of real substrates.
The effect of $\V$ in our model is to shift the chemical potential $\mu$ of the
gas (see Sec.~\ref{sec:V_shifts_chem_pot}). Because our ideal substrate shifts
the chemical potential of the gas molecules by providing a spatially uniform
potential energy field, the entropy of the gas in the ideal substrate is equal
to the entropy of the gas in its bulk state at the same density and
temperature. In contrast, a real substrate provides a non-spatially uniform
potential. Consequently, the entropy of the gas inside a real substrate is
\emph{not} equal to the entropy of the bulk gas at the same density and
temperature. Therefore, the parameter analogous to $\V$ in a real substrate
will involve both energy and entropy. The real-substrate analog of $\V$ is the
shift of molar Gibbs free energy provided by the substrate, specifically an
\emph{isosteric} (or constant-density) shift of the Gibbs free energy:

\begin{equation}
   \gst(\rho, T) \equiv
    \mu_{\text{ads}}(\rho, T) - \mu_{\text{gas}}(\rho, T).
  \label{eq:g_st}
\end{equation}
The isosteric Gibbs free energy difference $\gst$ is the difference in molar
Gibbs free energy (equivalent to chemical potential) between the adsorbed gas system and
the bulk gas \emph{with the same density of gas molecules} and at the same temperature. The quantity $\gst$
does \emph{not} correspond to a change in the molar Gibbs free energy as a
molecule is adsorbed, which is zero under conditions of coexistence. The
quantity $\gst$ in a real substrate is a direct analog to $\V$ in our ideal
substrate because it is the chemical potential shift needed to impose on the
bulk gas in coexistence with the real substrate to achieve the same density as
in the substrate (compare with eqn.~\ref{eq:xi_vs_xi0}).
%
%
Figure~\ref{fig:delta-G-cartoon} illustrates a hypothetical experiment to
measure $\gst$.  Two containers of equal volume are connected by a hose.  One
container holds the substrate, and the other is empty (apart from gas).    The
altitude of one container is adjusted until the number of gas molecules is equal
in each container.  The value of $\gst$ is the gravitational potential energy
difference of gas molecules in the two boxes.

Note that $\gst$ is a property
of both the substrate and the identity of the gas. Because real substrates
offer a \emph{non}-spatially uniform potential, $\gst(\rho, T)$ is a function
of $\rho$ and $T$, unlike our ideal, homogenous substrate where $\gst(\rho,
T)=\V$. Consequently, throughout this article, we show $\gst(\rho, T)$ for real
substrates at both conditions relevant to gas storage and delivery, $\pfull$
and $\pempty$.

In practice, we can readily compute $\gst(\rho, T)$ of a real gas/substrate
system from (i) the (experimental or simulated) equilibrium adsorption isotherm
of the gas in the substrate and (ii) the (experimental or simulated) chemical
potential of the bulk gas. Consider the real substrate in thermodynamic
equilibrium with a bulk gas at fixed temperature $T$ and pressure $p$, and let
$\rho=\rho(p, T)$ be the density of gas in the substrate. At coexistence, the
chemical potential of the bulk gas is equal to the chemical potential of the
adsorbed gas in the substrate. Thus, we can use the experimentally known molar
Gibbs free energy of the pure, bulk gas system at temperature $T$ and pressure
$p$ to determine the molar Gibbs free energy of the adsorbed system:
$\mu_{\text{ads}}(\rho, T)=\mu_{\text{gas}}(p, T)$. We can then also look up
the known chemical potential of the bulk gas at the same density and
temperature as in the substrate, $\mu_{\text{gas}}(\rho, T)$. Via
eqn.~\ref{eq:g_st}, $\gst$ follows from subtracting the two quantities.

\begin{figure}
    \centering
    \includegraphics[width=0.95\columnwidth]{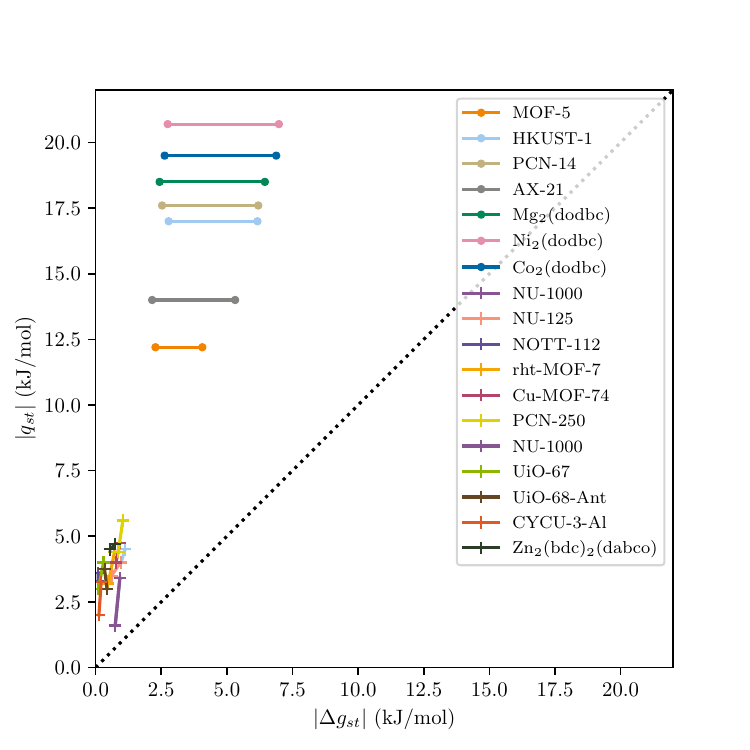}
    \caption{Relationship between $\gst$ and the isosteric heat
      $q_{st}$ for several prominent adsorbents at room
      temperature (data from Ref.~\cite{mason2014evaluating, garcia2018benchmark}). The dots represent the properties of methane
      adsorption at 5.8~bar and 65~bar. The $+$'s represent the
      properties of hydrogen adsorption at 5~bar and
      100~bar.}
    \label{fig:qst-vs-delta-G}
\end{figure}

An interesting question is how $\gst$ relates to the commonly measured and
reported isosteric heat of adsorption $q_{st}$, which is roughly the energy
change when a gas molecule is adsorbed~\cite{sircar1999isosteric,
tian2017differential}. Figure~\ref{fig:qst-vs-delta-G} shows how $q_{st}$
compares to $\gst$ for several prominent adsorbents. In every case,
$|q_{st}|>|\gst|$ because the gas in the adsorbent always has less entropy than
the gas in the bulk at the same density and temperature. That is, while
adsorption is energetically favored, it is entropically disfavored due to the
restrictions imposed on the configuration of the gas molecules via steric
interactions with the substrate itself; this counters the energetic attraction.

\begin{figure}
    \centering
    \includegraphics[width=0.95\columnwidth]{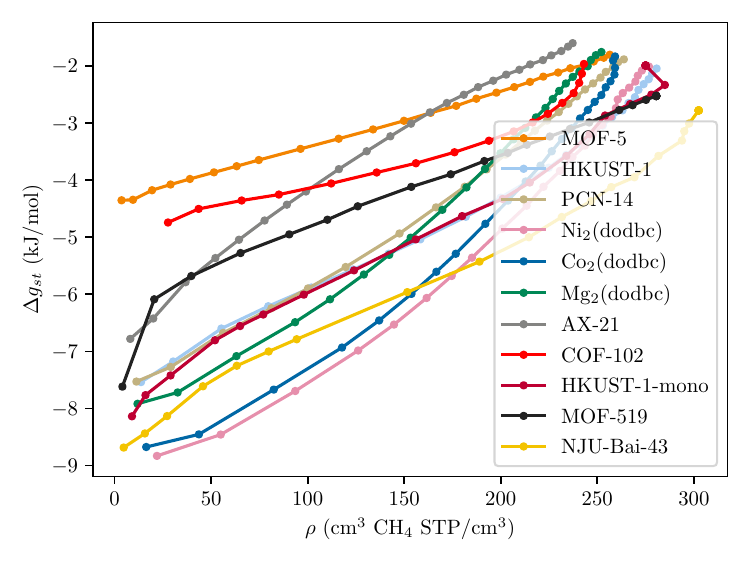}
    \caption{The density-dependence of $\gst$ of methane in several
      adsorbents (298\ K). Note that $\gst$ is monotonic in
      $\rho$. Data from
      Refs.~\cite{mason2014evaluating, furukawa2009storage}.
    }
    \label{fig:methane-gst}
\end{figure}

\subsection{An upper bound when $\gst(\rho)$ is monotonic}\label{sec:monotonic}
Every adsorbent has a $\Delta g_\text{st}$ at $\pfull$ and $\pempty$ corresponding
to a full and empty density, $\rhofull$ and $\rhoempty$, respectively. The deliverable capacity of the adsorbent is equal to
the difference between the full and empty density. Examining
Fig.~\ref{fig:methane-gst}, we see that a significant variety of known
experimental $\Delta g_\text{st}$ curves are monotonic. Furthermore, our
qualitative argument in Section~\ref{sec:upper-bound} suggests that this
function \emph{should} monotonically increase for rigid substrates, as
increasing the density of gas causes some of the gas to reside in higher-energy
sites. If this is always the case, the deliverable capacity of a real material
(with a non-spatially uniform potential) with a given $\gst$ at $\pfull$ or
$\pempty$ is bounded below by the deliverable capacity in our idealized
substrate with $\V=\gst$, even for a value of $\V$ that is not optimal.

\begin{figure}
    \centering
    \includegraphics[width=0.95\columnwidth]{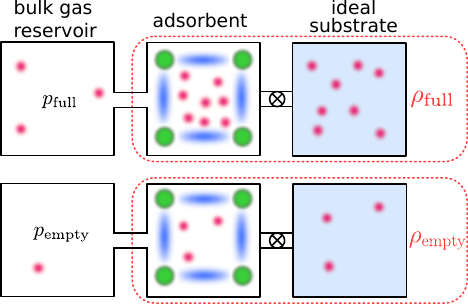}
    \caption{A bulk gas reservoir is connected to a volume filled with adsorbent material (green/blue: material atoms, red: gas particles). The adsorbent is also connected, but with a closed valve, to an equal volume containing the ideal substrate (shaded blue background represents the uniform potential energy field $\V$) with an equal density of gas as in the adsorbent. The top (bottom) shows a realization of this connected system where the pressure of the bulk gas is $\pfull$ ($\pempty$), and thus the density of gas in the adsorbent is $\rhofull$ ($\rhoempty$) by definition. The deliverable capacity of the adsorbent is $\rhofull - \rhoempty$. We can determine if the deliverable capacity of gas in the ideal substrate is lower or higher than in the adsorbent by considering what happens when we open the valves connecting the two pairs of volumes containing adsorbent and the ideal substrate.}
    \label{fig:delta-gst-maximum}
\end{figure}

To show this, we perform a thought experiment illustrated in
Fig.~\ref{fig:delta-gst-maximum}. We begin with two pairs of volumes.  Two of
these volumes contain a real adsorbent material.  These adsorbent volumes are
connected to bulk gas reservoirs at $\pfull$ and $\pempty$ respectively, and
by definition contain gas at density $\rhofull$ and $\rhoempty$.  The other two volumes
volumes contain gas at a density equal to the two adsorbent volumes ($\rhofull$ and $\rhoempty$), but contain our
idealized substrate with uniform potential energy $\V$.

We now consider what happens if we open a diffusive connection between a
volume of adsorbent and a volume with an idealized substrate that initially contain
the same density of gas, for instance by connecting them with a tube. If the
chemical potential in the idealized substrate is higher than in the porous material,
then gas will flow out of the idealized substrate, lowering its density.  Conversely,
if the chemical potential is lower in the idealized substrate than in the
volume of porous material, then gas will flow into the idealized substrate, increasing
its density.
The difference in chemical potential between those two volumes is
\begin{align}
   \gst(\rho) - \V &= \left(\mu_{\text{ads}}(\rho) - \mu_{\text{gas}}(\rho)\right)
   - \left(\mu_{\V}(\rho) - \mu_{\text{gas}}(\rho)\right)
   \\
   &= \mu_{\text{ads}}(\rho) - \mu_{\V}(\rho).
\end{align}
where $\mu_{\text{ads}}(\rho)$ and $\mu_{\V}(\rho)$ are the chemical potentials
of gas in the porous material and ideal substrate with potential $\V$, respectively.
Thus if $\gst(\rhoempty) =
\V$, the low-density containers will remain at their initial density after
they are connected. Thus, the deliverable capacity of the adsorbent will be greater
than the deliverable capacity of the ideal substrate with uniform potential $\V$ if and
only if $\gst(\rhofull)<\V$, i.e. if $\gst(\rho)$ does \emph{not} monotonically
increase, since then the gas in the high-density system will flow from the ideal
substrate to the adsorbent. By the same token, if we consider the case where
$\gst(\rhofull) = \V$, then for the adsorbent to achieve a greater deliverable
capacity than the idealized substrate, the adsorbent must have less residual gas,
which means that gas must spontaneously flow from the low-density adsorbent to the
volume with potential $\V$, which means that $\gst(\rhoempty)>\V$. Once again,
exceeding our upper bound requires a material with a $\gst(\rho)$ that does not
increase monotonically.

Taken together, this indicates that not only is our absolute upper bound an
upper bound for rigid adsorbents, but the green curve labeled $\rho_D(\gst)$ in
Figs.~\ref{fig:methane-298-D}, \ref{fig:hydrogen-298-D}, and
\ref{fig:hydrogen-77-D} is an upper bound for materials with a non-optimal
$\gst$ at either $\pfull$ or $\pempty$.


\section{Results}
Although our theoretical framework allows us to place an upper bound on the
deliverable capacity of many different gases in rigid porous materials, we
focus on the maximal deliverable capacity of methane and hydrogen gas in our
homogeneous substrate due to their application as transportation fuels. To
characterize the density of the gases, $\rho_g(p, T)$, we use experimental data
from NIST~\cite{nist}, which naturally includes quantum effects that
particularly affect hydrogen at low temperature~\cite{kumar2006quantum}. In the
context of storage onboard passenger vehicles, we compare our upper bound with
several prominent porous materials using experimental adsorption isotherms from
the literature; we also compare with deliverable capacity targets set by the US
DOE.

Figure~\ref{fig:methane-298-D} shows the upper bound for methane storage at
room temperature~(374~cm$^3$~STP/cm$^3$). In addition to the predicted maximum
deliverable capacity as a function of $\V$, the ARPA-E target of
315~cm$^3$~STP/cm$^3$~\cite{arpaemove} is shown for context. For an adsorbent
with at least 84\% void fraction, the ARPA-E target is theoretically possible.
The experimental deliverable capacities of several
adsorbents~\cite{mason2014evaluating} are also shown over a range of $\gst$
(converted from the measured adsorption isotherm as explained in
Sec.~\ref{sec:phi-is-delta-g}). For context, the highest observed
deliverable capacity for methane at room temperature in a rigid material is 208~cm$^3$~STP/cm$^3$~\cite{simon2015materials}.

\begin{figure}
    \centering
    \includegraphics[width=0.95\columnwidth]{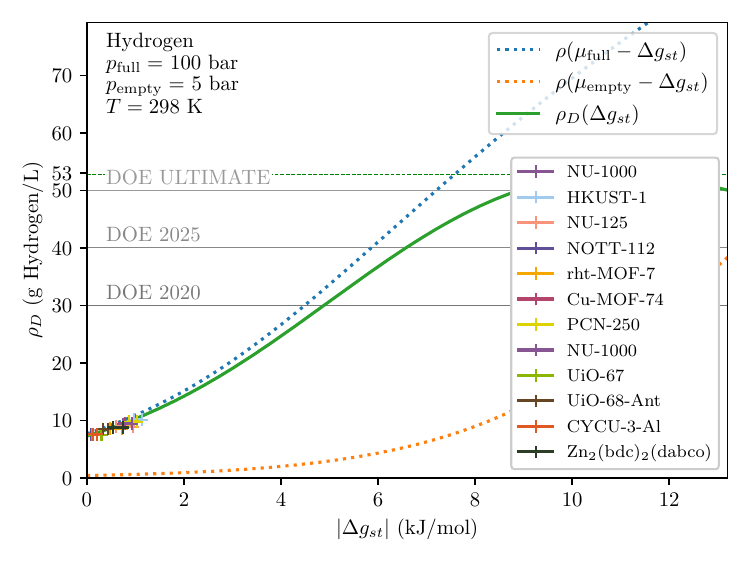}
    \caption{Deliverable capacity of hydrogen at room temperature as a function of the attractive Gibbs free energy $|\gst|$.  Experimental deliverable capacities for several porous materials (data from Ref.~\cite{garcia2018benchmark}) are shown along with the experimental values for $\gst$ at the empty and full pressures shown as $+$'s connected by a line.}
    \label{fig:hydrogen-298-D}
\end{figure}

\begin{figure}
    \centering
    \includegraphics[width=0.95\columnwidth]{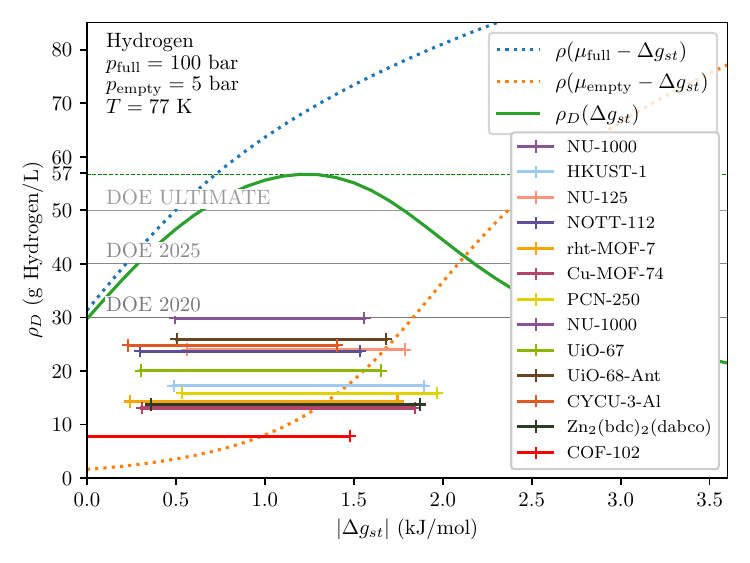}
    \caption{Deliverable capacity of hydrogen at 77\ K as a function of the attractive Gibbs free energy $\gst$. Experimental deliverable capacities for several porous materials (data from Ref.~\cite{garcia2018benchmark, furukawa2009storage}) are shown along with the experimental values for $\gst$ at the empty and full pressures shown as $+$'s connected by a line.}
    \label{fig:hydrogen-77-D}
\end{figure}

Storage of hydrogen is considerably more challenging owing to its relatively
weak interaction with adsorbents. For room temperature storage, the DOE
ULTIMATE deliverable capacity target~\cite{DOE} is within 6\% of the upper
bound. Figure~\ref{fig:hydrogen-298-D} shows the theoretical upper bound curve
for hydrogen storage along with experimental measurements for known adsorbents. The
DOE ULTIMATE deliverable capacity target is theoretically possible, however by
such a small margin that we can safely rule out the possibility of reaching
this target through storage and delivery of hydrogen in \emph{any} rigid
substrate at room temperature. Such a material would require a void fraction of
at least 94\%. On top of this, meeting the DOE ULTIMATE target requires an optimal
$|\gst|$ of 10~kJ/mol which is far greater than what is found in observed
porous materials. This reflects the known fact that hydrogen interacts with
substrates far more weakly than methane does. 
One approach to increase the deliverable capacity is to reduce the storage
temperature. This is illustrated in Fig.~\ref{fig:hydrogen-77-D}, which shows
the upper bound to the deliverable capacity of hydrogen at 77\ K, the boiling
point of nitrogen. The DOE ULTIMATE target in this case looks far more
achievable, and with a much lower $|\gst|$. In fact, an empty tank at this
temperature can satisfy the DOE 2020 target. The DOE ULTIMATE target is 14\%
below the upper bound. Actual adsorbents fall far short of the theoretical maximum.

Why do we not experimentally observe materials approaching the upper bound on
the deliverable capacity? First and foremost, any adsorbent substrate will be composed of
atoms, which will exclude the gas from some volume. Due to strong short-range
interactions, from a longer length-scale perspective, substrate atoms must be approximately uniformly distributed to achieve strong
attraction for gas atoms, imposing a limit on the pore
volume. Second, in contrast to the homogeneous substrate, real materials have a spatially non-uniform attraction for
gas atoms. As a result, there are regions of space with
non-optimal attraction. For hydrogen, there is the further issue that there are
no known \emph{physical} interactions that are sufficiently strong to give an
optimal deliverable capacity at room temperature.

\begin{figure}
    \centering
    \includegraphics[width=0.95\columnwidth]{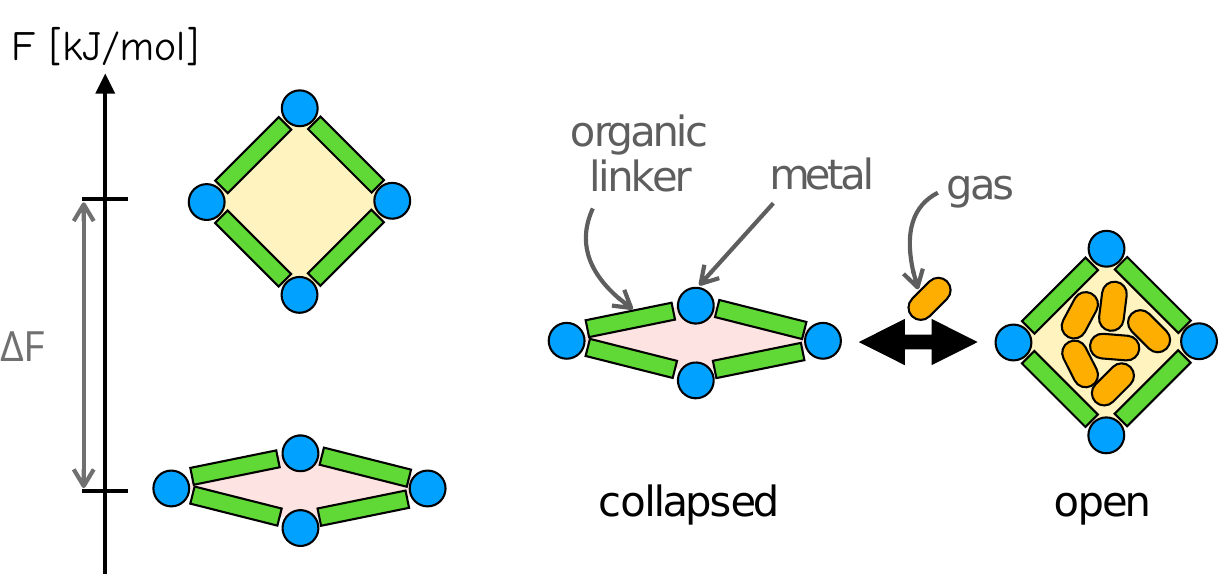}
    \caption{Cartoon of a two-phase, collapsing material. The material can exist in two phases: an open phase in which gas molecules can adsorb, and a closed phase that is collapsed and cannot fit any gas molecules. Take the closed phase as more stable, so that the material can collapse and expel residual gas.
    The pressure of gas at which the structural transition occurs depends on the gas-host interactions and relative stability of the phases. We wish for the structural transition to occur between $\pfull$ and $\pempty$.
    }
    \label{fig:flexible-cartoon}
\end{figure}

\section{Flexible two-phase model}
\begin{figure}
    \centering
    \includegraphics[width=0.95\columnwidth]{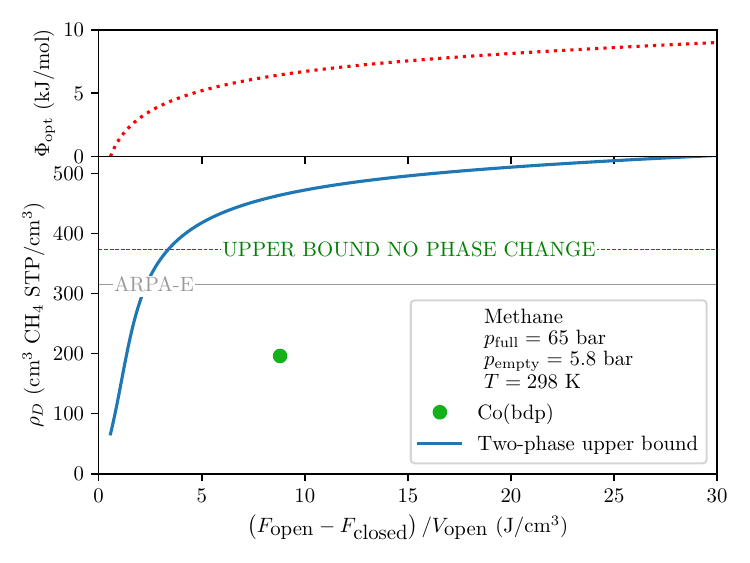}
    \caption{Maximal possible deliverable capacity of methane in a two-phase, collapsing substrate at 298\ K as a function of the free energy difference 
      between open and closed phases. The experimental deliverable capacity for Co(bdp) is shown for
       comparison, along with the free energy difference between open and closed conformations~\cite{mason2015methane}.}
    \label{fig:methane-flexible}
\end{figure}
While our deliverable capacity limit does not apply to materials that change
conformation in response to gas adsorption, here we adapt our theoretical
framework to pertain to bistable adsorbents modeled after Co(bdp)
\cite{mason2015methane}. Actuated by the presence of gas, Co(bdp) can switch
between two distinct structural phases: an ``open'' phase where gas molecules
can fit in the pore space and a collapsed, ``closed'' phase in which no gas can
fit inside.

We follow Coudert et al. \cite{coudert2008thermodynamics} in developing a
thermodynamic model for a two-phase, collapsing adsorbent like Co(bdp). 
Let $\Delta F=F_{\text{open}}-F_{\text{closed}}$ be the Helmholtz free energy difference per
mole of adsorbent between the two phases in the absence of gas. Suppose the
closed phase is more stable ($\implies \Delta F >0$), so that upon a reduction
in the pressure of the gas, the material can collapse and expel all residual
gas to enhance the deliverable capacity compared to rigid materials that will
retain some gas at $\pempty$. In addition, assume the open phase hosts a
uniform, attractive potential energy field $\V<0$ for gas molecules to adsorb,
so that eqn.~\ref{eq:xi_vs_xi0} holds for the material when in the open phase.

This model of a two-phase, collapsing material has no upper bound on the
deliverable capacity, since $|\V|$ and $\Delta F$ can together be increased
without bound.
We can, however, place a bound on the deliverable capacity for a fixed $\Delta F$, since if
the gas is too strongly attracted to the open phase, the material will remain
open at $\pempty$, and, then, our rigid upper bound applies.
We place an upper bound on the deliverable capacity in the two-phase material by identifying the
attractive interaction $\V$ that balances the gas-host attractions in the
idealized-open-substrate at $\pempty$ with the free energy penalty to open.

For an enhancement of the deliverable capacity by this two-phase, flexible material, 
the open phase must be preferred at $\pfull$, with the closed phase preferred at $\pempty$. 
The upper bound for the two-phase material is found by identifying the largest
possible attraction $|\V|$ consistent with the material transitioning between
$\pempty$ and $\pfull$ to expel all residual gas.
The material changes phases when the osmotic potentials of the open and closed systems 
with and without gas, respectively, are equal \cite{coudert2008thermodynamics} (neglecting any hystersis).
Thus, the optimal background potential $\V_{opt}$ equates the grand potential of the gas in the open phase 
with the free energy difference between the two phases at $\pempty$:
%
\begin{align}
    \frac{\Delta F}{V_{\text{open}}} &= \int_{-\infty}^{\muempty} \rho_g(\mu-\V_{opt})d\mu, \label{eq:v_opt_bistable}
\end{align}
where we follow Coudert \emph{et al.}\cite{coudert2008thermodynamics} and
neglect the $pV$ terms, and $V_{\text{open}}$ is the molar volume of the open
phase.
I.e., $|\V_{opt}|$ gives the maximal gas-open-host attractions under the constraint that the material collapses and expels all residual gas at $\pempty$; if the attraction were greater, the material would refuse to collapse and expel its gas at $\pfull$.
The deliverable capacity is then the density of gas in the open phase at $\pfull$, $\rho_g(\mufull{} - \V_{opt}, T)$ with $\V_{opt}$ satisfying eqn.~\ref{eq:v_opt_bistable}.

\begin{figure}
    \centering
    \includegraphics[width=0.95\columnwidth]{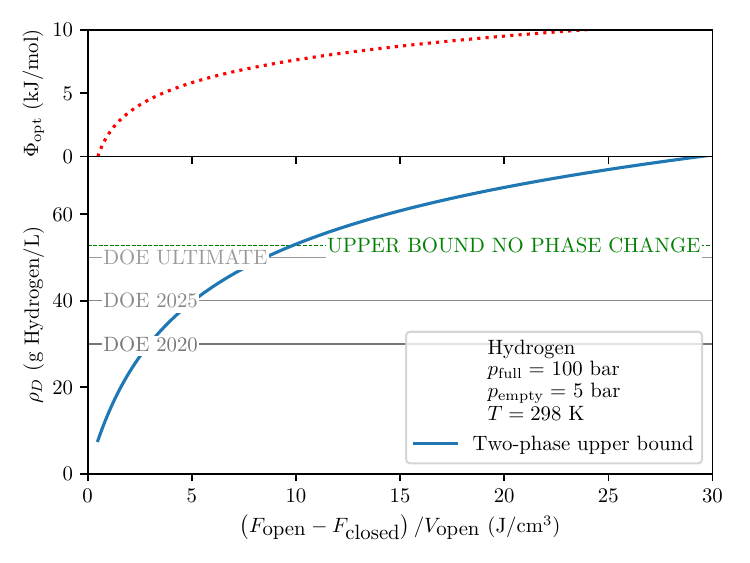}
    \caption{Deliverable capacity of hydrogen in a flexible, two-phase substrate at 298\ K as a function of the free energy difference 
      between open and closed phases.}
    \label{fig:hydrogen-flexible}
\end{figure}

Figure~\ref{fig:methane-flexible} illustrates
how the upper bound on the deliverable capacity of methane in a two-phase,
collapsing material depends on $\Delta F$-- and the $\V_{opt}$ that maximizes the deliverable capacity. 
For comparison, we include the free energy
difference for Co(bdp)~\cite{choi2008broadly} and its measured deliverable capacity.
Indeed, the deliverable capacity of a two-phase, collapsing material can exceed that of a rigid material if 
(i) the closed phase is sufficiently stable relative to the open phase
and
(ii) the gas has a correspondingly high affinity for the open phase.
Figure~\ref{fig:hydrogen-flexible} shows the corresponding result for hydrogen storage, indicating that, for a flexible material to
reach the DOE ULTIMATE deliverable capacity, it would require far greater stability for the closed
phase, as well as an unprecedented level of attractive interaction for hydrogen.

\section{Conclusions}
We have established an upper bound on the deliverable capacity via
pressure-swing adsorption in rigid porous solids based on experimentally
measured properties of pure gases. While these upper bounds do not rule out the
discovery of materials that reach current DOE targets for room temperature hydrogen and methane storage, they cast strong doubt
on the possibility of achieving these goals when we consider the additional
constraints imposed due to steric hindrance between substrate atoms and the
adsorbate. Our upper bound does indicate that those goals cannot be exceeded by
more than 16\% for methane and 6\% for hydrogen. Fortunately, there are some
limitations to these upper bounds which suggest avenues for future developments.

The first limitation of our proof is that we restricted ourselves to
\emph{isothermal} pressure swing storage (which is consistent with the targets set by DOE). By raising the temperature of the
adsorbent during gas discharge to drive off residual
gas~\cite{gomez2014exploring}, the deliverable capacity could be enhanced,
albeit at the cost of a more complicated engineering design of the fuel tank
and vehicle.
We cannot place an upper bound on such a process using our approach because, with
changing temperature, a uniform potential may not give the maximum deliverable
capacity.  A nonuniform potential will lower the entropy of the adsorbed gas,
which will make the gas easier to release by raising the temperature.

The second limitation of our proof arises in the assumption of a rigid
substrate. The rigid substrate acts as a static potential energy field for gas
molecules that is unchanged by the adsorption of gas. For most porous materials
this is a reasonable approximation, and this assumption is frequently made in
both the simulation and theory of porous materials~\cite{duren2009using}.
However, there are cases where the substrate can provide a very strong
gas-density-dependent interaction through structural
flexibility~\cite{schneemann2014flexible}. A flagship example is MOF
Co(bdp)~\cite{choi2008broadly}, which possesses a wine-rack-like topology
capable of hinge motion. At low methane pressure, Co(bdp) adopts a collapsed,
nonporous state, but expands to a porous state and fills with gas at higher
pressures~\cite{mason2015methane}. This allows Co(bdp) to fully expel its
residual gas at low pressures. Flexible materials could have significantly 
higher deliverable capacities, as we have shown for two-phase structures 
that collapse and expel all residual gas at low pressures.

Several studies elucidated relationships between structural features of
nanoporous materials---such as void fraction, pore size, and surface area---and
their deliverable capacity of various gases
\cite{gomez2014exploring,ahmed2019exceptional,simon2015materials,moghadam2018computer,Tong2018}.
Such structure-property relationships serve as useful heuristics for designing
new materials with high deliverable capacity. However, the intention of our
paper is not to discover structure-property relationships. Rather, we place our
upper bound on the deliverable capacity of [rigid] materials that holds
irrespective of how the structural features of the material are tuned.



For vehicular gas storage, gravimetric deliverable capacity (amount of gas per
mass of material) may be considered as a material performance metric to avoid a
heavy adsorbed-gas fuel tank. However, the volumetric deliverable capacity
(amount of gas per volume of material) dominantly determines the driving range
on a full fuel tank of a given volume \cite{mason2014evaluating}. That said,
several articles have explored the tradeoff between gravimetric and volumetric
deliverable capacity, particularly for hydrogen storage
\cite{gomez2017understanding,chen2020balancing,goldsmith2013theoretical,ahmed2021predicting}.


We note that our upper bound can readily be applied to the storage of other
gasses of interest, with the proviso that the gas is far from crystallization.
Our code is available at \href{https://https://github.com/droundy/thesis-pommerenck/tree/master/gas-adsorption}{https://https://github.com/droundy/thesis-pommerenck/tree/master/gas-adsorption}.


\bibliography{bibfile} 

\end{document}